\documentclass[aps,pra,twocolumn,tightenlines,longbibliography,superscriptaddress,doublespace,nofootinbib]{revtex4-2} 

\usepackage{graphicx}
\usepackage{setspace,color,appendix,physics,wasysym,xcolor,xparse}
\usepackage{appendix} 
\usepackage[english]{babel}
\usepackage{amsmath,amssymb,amsfonts,mathtools}
\usepackage{mathtools}
\definecolor{brickred}{rgb}{0.8, 0.25, 0.33}
\definecolor{celestialblue}{rgb}{0.29, 0.59, 0.82}
\definecolor{cornflowerblue}{rgb}{0.39, 0.58, 0.93}
\definecolor{denim}{rgb}{0.08, 0.38, 0.74}
\definecolor{armygreen}{rgb}{0.29, 0.33, 0.13}
\definecolor{cardinal}{rgb}{0.77, 0.12, 0.23}
\definecolor{carnelian}{rgb}{0.7, 0.11, 0.11}
\definecolor{armygreen}{rgb}{0.29, 0.33, 0.13}

\usepackage[colorlinks=true, urlcolor=black, citecolor=blue, linkcolor=red, citebordercolor={1 0 0}, linkbordercolor={0 0 1}]{hyperref}

\usepackage{verbatim,multirow}
\usepackage[T1]{fontenc}
\usepackage{lmodern}
\usepackage{pxfonts}

\newcommand\identity{1\kern-0.25em\text{l}}

\begin{document}

\title{Energy efficient optical tracking for space quantum communication}

\author{Eric Vokes}
\affiliation{Department of Computer Science, University of York, York YO10 5GH, U.K.}
\author{Vinod N. Rao}
\affiliation{School of Physics, Engineering \& Technology, University of York, YO10 5FT York, U.K.}
\author{Elinore Spencer}
\affiliation{School of Physics, Engineering \& Technology, University of York, YO10 5FT York, U.K.}
\author{Rupesh Kumar}
\email{rupesh.kumar@york.ac.uk}
\affiliation{School of Physics, Engineering \& Technology, University of York, YO10 5FT York, U.K.}

\begin{abstract}
Power consumption is a critical constraint for CubeSat based quantum communication, where tracking systems often dominate the onboard power budget. We demonstrate an energy-efficient approach that enables reliable satellite tracking at substantially reduced beacon power by treating tracking as a weak-signal estimation task. Using a closed-loop system with fine steering mirrors and higher-order Kalman filters on ground, we can maintain stable tracking at a transmitted power equivalent to 34 mW over a -60 dB satellite to ground optical channel. Our results show that the resulting penalties on QKD bit error rates and signal-to-noise ratios are negligible, allowing for more efficient power allocation to quantum payloads in CubeSat missions.
\end{abstract}

\maketitle

\section{Introduction \label{sec:intro}}

Quantum communication enables two distant parties to share a secret key, which are proven to be information theoretic in nature \cite{scarani2009security, xu2020secure, portmann2022security}. Satellite based quantum communication offers a route to globally secure key distribution, overcoming the exponential attenuation limitations of terrestrial optical fibre links \cite{yin2017satellite, liao2017satellite, sidhu2021advances}. Flagship missions such as Micius \cite{lu2022micius}, Jinan-1 \cite{li2025microsatellite} have successfully demonstrated space-to-ground entanglement distribution and quantum key distribution (QKD) at intercontinental distances. At the CubeSat scale, experiments like SOCRATES \cite{takenaka2017satellite}, SpooQy-1 \cite{villar2020entanglement} and planned missions like QUBE \cite{knips2022qube}, SPOQC \cite{SPOQC}, SpeQtre \cite{Speqtre}, QEYSSat \cite{scott2020qeyssat}, etc \cite{armengol2008quantum, jennewein2014nanoqey, balakier2025high} are extending these capabilities into cost effective platforms. A demonstration of QKD from CubeSat may require the platform to be of volume 3U to 12U, depending mission objectives. However, a major portion of the space is occupied by the beam delivery and PAT system and other functional components for the satellites other than the QKD payload \cite{fitzpatrick2025speqtre}.

A representative CubeSat architecture for demonstrating downlink-based QKD typically comprises three main functional segments \cite{crusan2019nasa, nieto2019cubesat, bouzoukis2025overview, SPOQC, cubesat101}: \textit{the quantum payload}, \textit{the optical terminal}, and \textit{the supporting subsystems}. \textit{The quantum payload} \cite{sagar2023design, mendes2024optical} operating at optical (or NIR) wavelength generates and manipulates quantum states of light and includes either a weak coherent pulse source or an entangled photon source for discrete variable (DV) QKD, while continuous variable (CV) QKD involves either amplitude and phase modulation of light from a laser or squeezed states of light. Single photon avalanche detectors are usually utilised in the case of DV based protocols (e.g., in uplink/downlink configuration or for the demonstration of entanglement in space), whereas the CV protocols involve coherent detection of its quadratures. \textit{The optical terminal} \cite{rodiger2020high, schmidt2022dlr}, along with a telescope of aperture around $8-20~\text{cm}$, incorporates a fine-pointing assembly and handles the downlink transmission of the QKD signals and uplink reception of the beacon laser sent from the optical ground station (OGS). Conventionally, the downlink beacon laser is sent through a smaller aperture ($2-5~\text{cm}$ optical terminal) for larger beam coverage on the ground. And for precision tracking purposes, the wavelength of the downlink beacon laser is well separated from the QKD wavelength. \textit{The supporting subsystems} \cite{cappelletti2020cubesat} include an attitude determination and control system (ADCS) composed of star trackers, gyroscopes, reaction wheels, and magnetometers for maintaining stable alignment; onboard data handling and control electronics; RF communication antennas for telemetry and key management; and power and thermal management systems consisting of solar panels, storage batteries, and active thermal regulation. All of these contribute to ensure stable operation and constant performance throughout orbital day-night cycles.

\begin{figure}[ht]
\centering
\includegraphics[width=\linewidth]{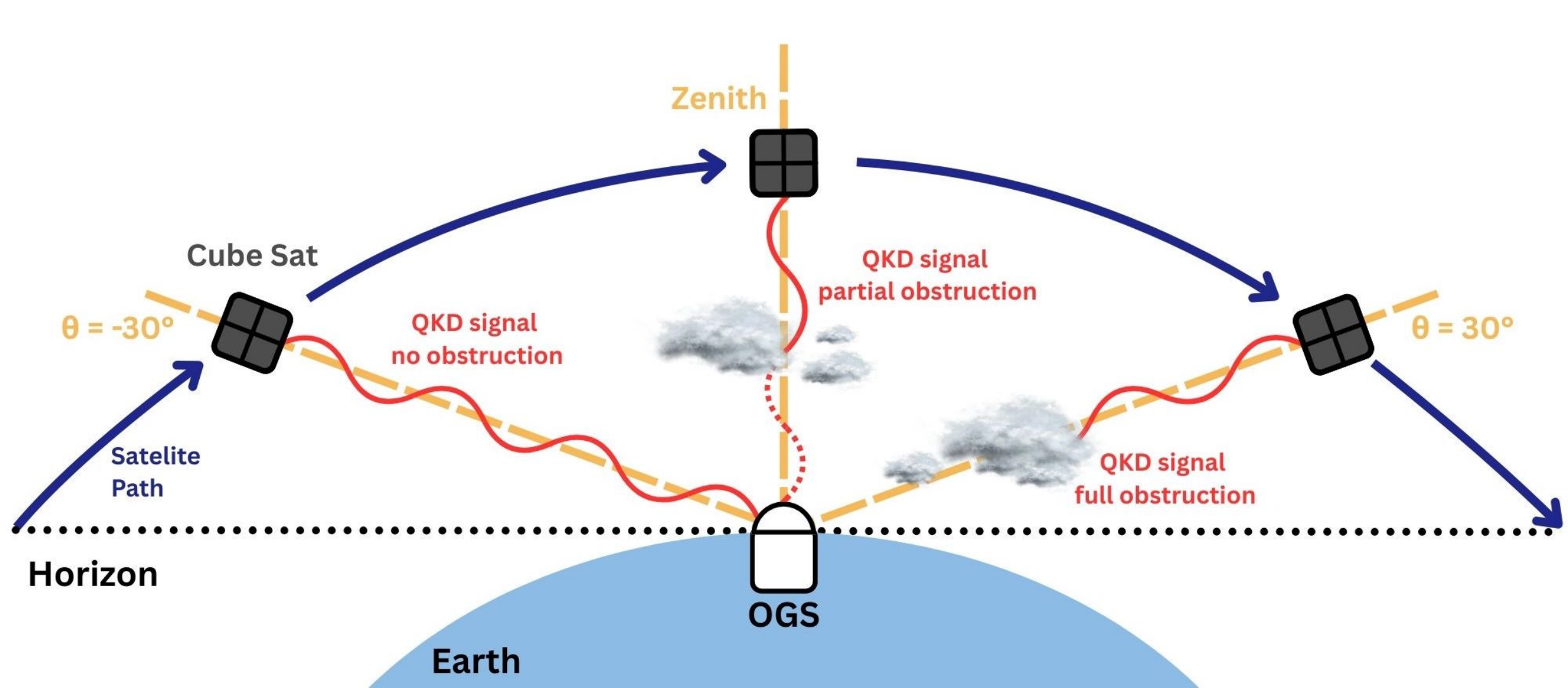}
\caption{Image depicting satellite pass over an OGS. Here the CubeSat involves the transmitter, that transmits signal to the OGS. The proposed QKD protocol involves tracking of CubeSat at a certain Zenith angle ($-30^\circ$), and then continue tracking until it reaches the last stage ($+30^\circ$).}
\label{fig:space}
\end{figure}

Power remains the dominant constraint in the case of a CubeSat. For a 3U CubeSat, body-mounted solar panels in low Earth orbit (LEO) yield an average electrical generation of approximately $5-7~\text{W}$, with higher peak power possible during sunlit periods but limited by eclipse duty cycles. More aggressive deployable solar arrays may temporarily boost instantaneous power, but typical configurations still average only $7-10~\text{W}$ in orbit \cite{santoni2014innovative, ibrahim2019comparison, yost2024state}. In contrast, a $20~\text{W}$ average is more characteristic of larger platforms, like 12 U CubeSats or small-satellite buses with extensive deployables \cite{shakoor2025comprehensive, yaqoob2022comprehensive, patel2023spacecraft}.

Optical quantum links require precise pointing accuracy because of their narrow beam divergence. Sub-milliradian pointing deviations can severely degrade link performance. A downlink beacon laser, and sometimes an uplink beacon, provide critical reference signals for acquisition and closed-loop fine-pointing stabilization \cite{giggenbach2017system}. Beam tracking typically employs two schemes:

\begin{itemize}
\item Open-loop tracking, which relies on orbital ephemeris and prediction to orient the transmitter or receiver;
\item Closed-loop tracking, which uses real-time feedback from a beacon-imaging camera to refine pointing and mitigate residual errors \cite{podmore2021qkd, giggenbach2023link}.
\end{itemize}

LEO downlinks suffer significant attenuations commonly $30-50~\text{dB}$ total loss due to diffraction, atmospheric absorption, turbulence, and misalignment \cite{lim2021centroid, giggenbach2023link}. For tracking cameras, this means that the beacon must exceed the noise floor within the exposure window to allow centroid detection. Photon-shot noise, detector noise, background atmospheric noise, and atmospheric scintillation set the minimum effective detection limit. The detection thresholds can range from a few tens to hundreds of photons per frame, depending on system parameters \cite{chu2021feasibility, li2023free, cai2024free, zhan2025long}.

Common CubeSat QKD concepts assume $4~\text{W}$ optical power for a downlink beacon laser with $50\%$ conversion efficiency requires $8-10~\text{W}$ electrical power \cite{roubal2025laser, barnes2007solid}, which can be a limiting factor for small CubeSat of size 6U and below as it constrains the power available for the quantum payload. QKD photon sources, detectors, and thermal controls already demand high electrical power, so power allocation must be carefully balanced. 

The presented work seeks to minimise the required beacon power by treating acquisition and tracking as a weak signal detection problem at the OGS. We analyse link loss and the tracking camera sensitivity to set detection thresholds and quantify how much beacon power can be reduced without degrading the closed-loop tracking stability. Specifically, we implement a Kalman filter to predict the satellite trajectory from the camera pixels, enabling fine tracking under changing acceleration as in Fig. \ref{fig:space}. We also demonstrate intermittent obstruction of the beacon laser by clouds does not affect the tracking. Finally, we try to estimate the penalty on lowering the beacon laser power in satellite tracking in terms of Quantum Bit Error Rate (QBER) in DV-QKD and Signal to Noise Ratio (SNR) in CV-QKD, both terms determine the secure key generation rate. We show that our approach allows more of the CubeSat power to be devoted to the quantum payload, improving the viability of QKD implementation on small satellites.

Below, we present our work where the necessity for high power beacon is relaxed in a CubeSat and this leads to the aspect of minimising the tracking error as well. After a brief introduction in Sec. \ref{sec:intro}, we provide the details of the methods in Sec. \ref{sec:met}. The results are given in Sec. \ref{sec:res} and finally we conclude in Sec. \ref{sec:conc}.

\section{Methods \label{sec:met}}

\begin{figure}[ht]
\centering
\includegraphics[width=\linewidth]{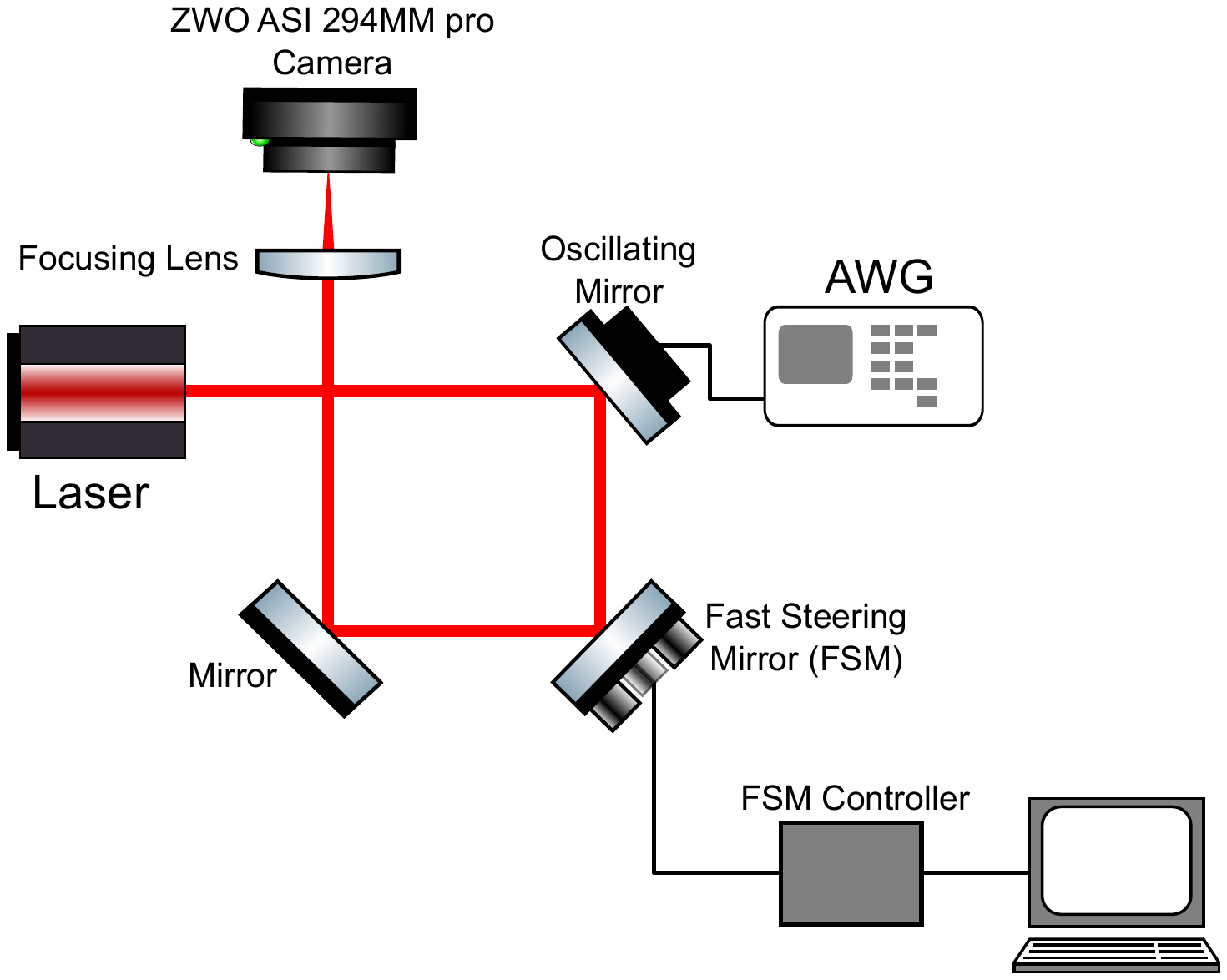}
\caption{Block diagram for emulating the tracking of satellite on a table top experiment. An oscillating mirror, driven by arbitrary waveform generator (AWG) displaces the beam where the FSM counteract the displacement. A computer connected to the camera acquires the frames, processes it and provides instructions to the FSM.}
\label{fig:block} 
\end{figure}

To test our proposal, we first constructed the ground segment of the beacon tracking system, as shown in Fig. \ref{fig:block}. The setup utilizes a collimated laser beam with a width of $2.27~\text{mm}$ at a $638~\text{nm}$ wavelength to represent the downlink beacon. This beam propagates toward a fine steering mirror (FSM) capable of a $\pm4^\circ$ angular deflection. The beam is then captured by a tracking camera (ZWO ASI294MM Pro) configured with a $2 \times 2$ binning. This provides an image size of $19.1 \times 13~\text{mm}$, a resolution of $4244 \times 2820$, and a pixel size of $4.63~\mu\text{m}$.
The FSM is driven via USB commands to produce linear beam deflections, while camera frames are captured at a rate of around 20 frames per second with the minimum $32~\mu\text{s}$ exposure time and $120$ unity gain. A focusing lens illuminates approximately $250,000$ pixels on the sensor. We set the maximum laser power to $5.55~\mu\text{W}$, which simulates the power received over a $60~\text{dB}$ loss optical channel. This value represents the typical loss at a $\pm 30^\circ$ elevation angle for a beam emerging from a $2~\text{cm}$ satellite aperture at a $700~\text{km}$ LEO altitude, assuming a ground telescope aperture between $40~\text{cm}$ and $60~\text{cm}$. This corresponds to an initial laser power of $5.6~\text{W}$ at the satellite. We have also added a oscillating mirror in the beam path to demonstrate beam displacement and subsequent correction by the FSM.

\subsection{beacon Identification \label{subsec:laser_id}}

\begin{figure}[ht]
\centering
\includegraphics[width=\linewidth]{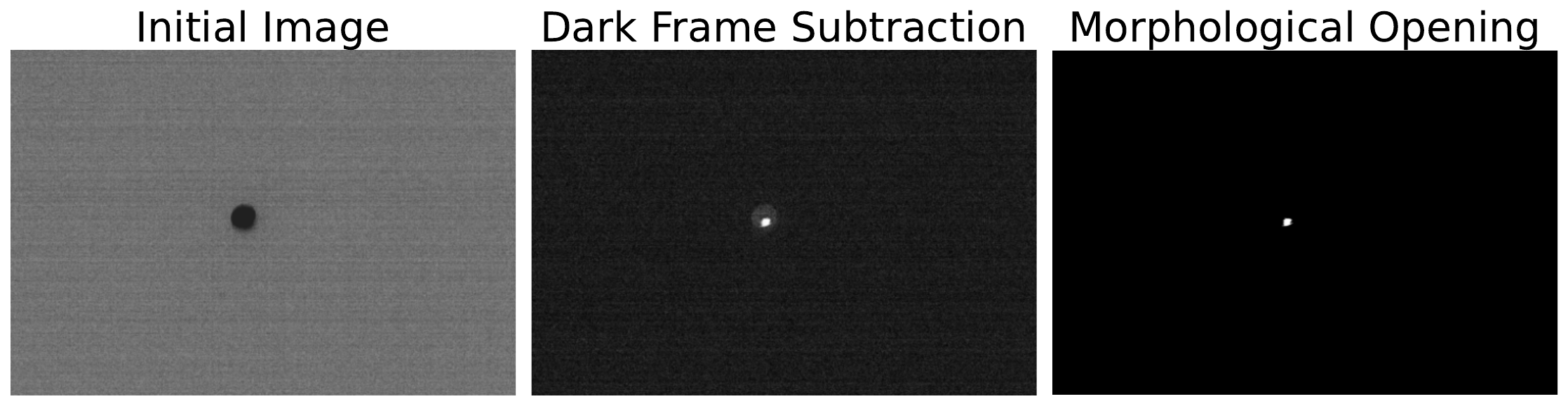}
\caption{Progression of the image pre-processing pipeline used to clean the camera frame. (a) The raw, unprocessed frame. (b) The frame after dark frame subtraction, removing fixed-pattern noise. (c) The final output after applying morphological opening, showing the isolated laser signal.}
\label{fig:cleaning}
\end{figure}

The captured frames are made to undergo the following processes with code written in Python with access to the OpenCV library. In order to differentiate and isolate the bright pixels due to the downlink beacon laser, from the dark pixels, the raw camera frame is cleaned in two steps. First, dark frame subtraction is applied to remove fixed pattern sensor noise, and although this removes the majority of artifacts, some stochastic salt-and-pepper noise will remain. This is then removed by applying morphological opening with a kernel size of 3 to erode small noise artifacts and then dilate the remaining foreground objects, as seen in Fig. \ref{fig:cleaning}.

To localise the beacon beam, a foreground mask is generated via Otsu's adaptive threshold \cite{otsu1975threshold}.
OpenCV’s contour detection is then utilised to isolate distinct shapes within the mask. Under the assumption that the largest detected contour represents the laser spot, image moments are calculated to determine the precise pixel coordinates of the beam's centroid. It is observed that at lower laser input power, the estimated beam area reduces to 20 percent of that of the maximum laser power, which results in non linear response graph given in the Fig. \ref{fig:calib_camera}.

\subsection{Camera calibration}
The pixel response of the camera to the input beacon laser needed to be calibrated to establish the minimum and maximum laser powers the camera sensor can handle. We assumed that an appropriate optical filter is used to filter out any unwanted light entering the camera sensor. The calibration was conducted in the absence of external light. Moreover, we consider that the satellite overpass is scheduled at night for maximising the QKD performance. Pixel intensity without the beacon laser is considered as the noise floor of the camera.

\begin{figure}[ht]
\centering
\includegraphics[width=\linewidth]{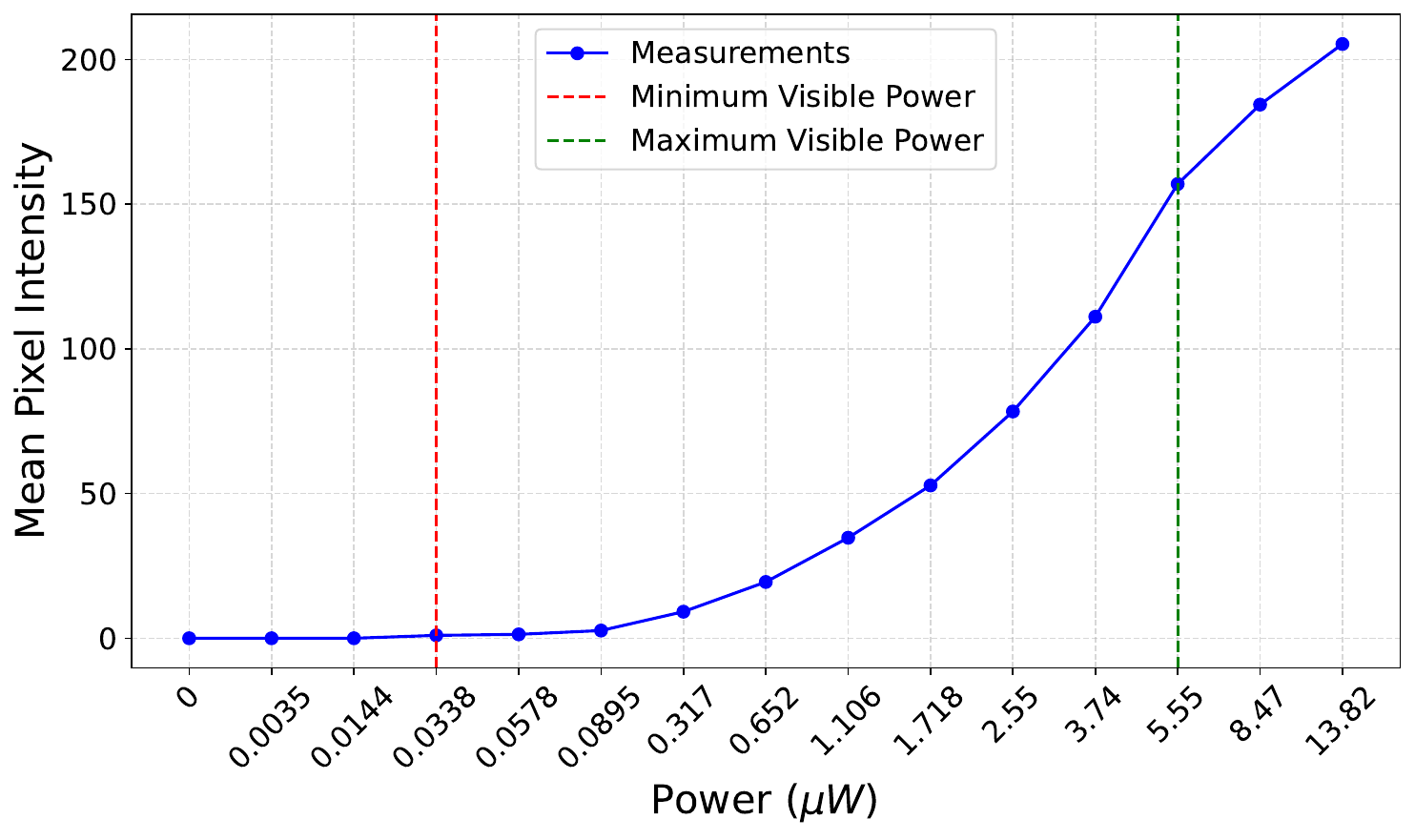}
\caption{The camera pixel mean intensity v/s laser beam power. The red and green dotted lines indicate the minimum and maximum laser intensities used during the tracking experiment, respectively.}
\label{fig:calib_camera}
\end{figure}

The Fig. \ref{fig:calib_camera} shows the camera pixel intensity response vs the input laser intensity. The non-linear response of the graph is due to variation in the estimation of the area of the beam at different beacon intensity levels while using Gaussian beam intensity profile. We have identified that the minimal intensity for our algorithm to identify the central pixel against the noise floor is around $0.03~\mu\text{W}$ which corresponds to $34~\text{mW}$ at the LEO satellite, over a $60~\text{dB}$ loss channel. The maximum intensity used is $5.55~\mu\text{W}$ which correspond to 5.6 W at the satellite.

\subsection{FSM calibration \label{subsec:fsm_nr}}

Before operation, an important calibration step must be performed to determine the distance ($d$) between the free space optics and the camera. This is done by actuating the FSM by a large known angle ($\theta_{\text{cal}}$) and then measuring the resulting laser displacement ($\Delta x$) from the camera frame. This distance was found to be $(\tan(\theta_{\text{cal}}) = \frac{\Delta x}{d})$. To ensure the laser beam remains centred to the cameras frame, a closed loop control algorithm was created between the camera frames and the FSM. During tracking, the calibrated distance is then used by the system to convert any measured pixel displacement into the precise angle required to recentre the beam until it overcomes the preset correction range of the FSM. At or beyond this stage, the telescopic mount will adjust the coarse tracking. 

\subsection{Emulating the satellite trajectory}

Though the pixel displacement can be caused by various reasons such as beam wandering; satellite jittering, vibration of the telescope, etc, here we consider it to be due to the motion of the satellite along its orbital trajectory. Stitching together the displacement values will give the satellite trajectory. However, in practice the FSM will counteract to nullify the displacement. Here we have considered (a) fixed pixel displacement account for a constant velocity ($\mathbb{CV}$) model and (b) variable pixel displacement account for constant jerk ($\mathbb{CJ}$) model of the trajectory. The required step wise displacement for emulating each model is known during the FSM calibration process.

The displacement values (i.e., the pixel index) serve as the input data for the Kalman filter for predicting the future pixel indices for both model (a) and (b). The later one represent the case of the realistic satellite trajectory as the relative velocity of the satellite with respect to the OGS is first accelerate from the horizon till the zenith angle and then decelerate toward the horizon.

\subsection{Kalman filter for constant velocity trajectory \label{subsec:cv_model}}

To establish a baseline for tracking performance, we initially implemented a linear Kalman filter based on a $\mathbb{CV}$ model. The $\mathbb{CV}$ assumes that the target moves with constant velocity in the image plane over each sampling interval, while any unmodeled acceleration is treated as process noise. The process starts with defining the system state, $\mathbf{x}_k$ at time $k$, consists of the beacon beam mean pixel positions, $x$ and $y$ and velocity $v_x$ and $v_y$ in the horizontal and vertical directions of the camera frame, written as $\mathbf{x}_k = [x, y, v_x, v_y]^T$. Then a linear state transition model evolves the state vector assuming constant velocity at the sampling interval $\Delta t$
\begin{equation}
\mathbf{x}_k= A_{\mathbb{CV}} \mathbf{x}_{k-1} + w_{k-1},
\end{equation}
where, the state propagation is governed by the transition matrix $A_{\mathbb{CV}}$:
\begin{equation}
A_{\mathbb{CV}} = \begin{bmatrix}
1 & 0 & \Delta t & 0\\
0 & 1 & 0 & \Delta t\\
0 & 0 & 1 & 0 \\
0 & 0 & 0 & 1
\end{bmatrix}.
\end{equation}

Here, $w_{k-1}$ is zero-mean Gaussian process noise that represents unmodeled disturbances. The prediction step of the $\mathbb{CV}$ Kalman filter is given by, 

\begin{align} 
\hat{\mathbf{x}}_{k \mid k-1} &= A_{\mathbb{CV}} \, \hat{\mathbf{x}}_{k-1 \mid k-1},\\
P_{k \mid k-1} &= A_{\mathbb{CV}} \, P_{k-1 \mid k-1} \, A_{\mathbb{CV}}^{T} + Q_{\mathbb{CV}},
\end{align}

where $\hat{\mathbf{x}}_{k \mid k-1}$ and $P_{k \mid k-1}$ denote the predicted state estimate and covariance at time $k$ based on measurements available up to time $k-1$. The covariance matrix with variances of position and velocity as the diagonal terms and position and velocity correlation on off diagonal terms.

\begin{equation}
P_k =
\begin{bmatrix}
\sigma_x^2 & 0 & \sigma_{x v_x} & 0 \\
0 & \sigma_y^2 & 0 & \sigma_{y v_y} \\
\sigma_{x v_x} & 0 & \sigma_{v_x}^2 & 0 \\
0 & \sigma_{y v_y} & 0 & \sigma_{v_y}^2
\end{bmatrix},
\end{equation}

And, $Q_{\mathbb{CV}} = \sigma^2_ {proc} \mathbf{I}_4 $ is the process noise matrix modeling noise with variance $\sigma^2_{proc}$.

Following the prediction, the actual measurement vector contains only position observations,
$\mathbf{z}_k = [x_k^{\text{meas}},\, y_k^{\text{meas}}]^T$ which is related to the state by $\mathbf{z}_k = H\text{x}_k + \mathbf{v}_k$, where 
\begin{equation}
H = \begin{bmatrix}
1 & 0 & 0 & 0 \\
0 & 1 & 0 & 0
\end{bmatrix},
\end{equation}
is the measurement matrix and $\mathbf{v}_k$ is the zero-mean Gaussian measurement noise.

The last step is the correction step in during which the Kalman gain is computed as
\begin{equation}
K_k = P_{k \mid k-1} H^{T} \left( H P_{k \mid k-1} H^{T} + R \right)^{-1},
\end{equation}
where $R = \sigma^2_{meas} \mathbf{I}_2 $ is the measurement noise covariance matrix with $\sigma_{meas}^2$ being the variance of measurement noise such as centroiding error, photon shot noise, detector read noise and background fluctuations, assumed to be identical along the $x$ and $y$ coordinates.

The state estimate is then updated according to
\begin{equation}
\hat{\mathbf{x}}_{k \mid k} = \hat{\mathbf{x}}_{k \mid k-1} + K_k \left( \mathbf{z}_k - H \hat{\mathbf{x}}_{k \mid k-1} \right),
\end{equation}

and the corresponding covariance update is given by
\begin{equation}
P_{k \mid k} = \left( I - K_k H \right) P_{k \mid k-1}.
\end{equation}

Both the process noise ($Q_{\mathbb{CV}}$) and measurement noise ($R$) covariances are empirically tuned with an automated offline grid search. The optimisation cost function minimized the root mean square (RMS) error between the estimated trajectory and the actual trajectory.

The $\mathbb{CV}$ model provides a simple and computationally efficient baseline for tracking. However, its accuracy degrades when a target exhibits significant acceleration, such as during satellite passes with rapidly changing line of sight dynamics. From the perspective of OGS, the satellite's motion is inherently curved and non-linear. Consequently, the Kalman filter must be updated to account for these accelerating trajectories.
 
\subsection{Kalman filter for an accelerating trajectory \label{subsec:cj_model}}

In the real case, with respect to the OGS, the satellite appears to move comparatively slower at the horizon and gradually accelerate towards the zenith and then decelerate. To accurately track and predict this more complex motion, a higher-order $\mathbb{CJ}$ model is introduced to analyse tracking performance under accelerating trajectories. This requires expanding the state vector to eight dimensions to include acceleration ($a_x,a_y$) and jerk ($j_x, j_y$) components, 

\begin{equation}
\mathbf{x}_k = [x, y, v_x, v_y, a_x, a_y, j_x, j_y]^T.
\end{equation}

Consequently, state transition matrix $A_{\mathbb{CJ}}$ is expanded to an $8 \times 8$ matrix that propagates the state forward using the integral equations of motion.
\begin{equation}
A_{\mathbb{CJ}} =
\begin{bmatrix}
1 & 0 & \Delta t & 0 & \tfrac{\Delta t^2}{2} & 0 & \tfrac{\Delta t^3}{6} & 0 \\
0 & 1 & 0 & \Delta t & 0 & \tfrac{\Delta t^2}{2} & 0 & \tfrac{\Delta t^3}{6} \\
0 & 0 & 1 & 0 & \Delta t & 0 & \tfrac{\Delta t^2}{2} & 0 \\
0 & 0 & 0 & 1 & 0 & \Delta t & 0 & \tfrac{\Delta t^2}{2} \\
0 & 0 & 0 & 0 & 1 & 0 & \Delta t & 0 \\
0 & 0 & 0 & 0 & 0 & 1 & 0 & \Delta t \\
0 & 0 & 0 & 0 & 0 & 0 & 1 & 0 \\
0 & 0 & 0 & 0 & 0 & 0 & 0 & 1
\end{bmatrix}.
\end{equation}

To minimise tracking error at the zenith the signs of the acceleration components ($a_x, a_y$) within the state estimate and the corresponding terms in the error covariance matrix are inverted. The other core matrices were also adapted to the 8-state model. The same tuning procedure described in Sec. \ref{subsec:cv_model} is then used to find the optimal values for the measurement noise covariance ($R$) and the process noise variance ($Q_{\mathbb{CJ}}$).

\subsection{Estimation of QBER and excess noise \label{subsec:qber}}

In order to evaluate the penalty for reducing the laser beacon power on the tracking, we can compare the relative increase in QBER in DV-QKD and SNR in CV-QKD. The tracking error reduces the signal coupling efficiency to the QKD receiver attached to the OGS. As a straightforward measure, we only need to estimate the relative increase in tracking error at lower beacon laser power compared to the normal case. Assuming the QKD signals follow a Gaussian beam profile, after the correction of turbulence-induced wavefront distortion by the adaptive optics system and Gaussian pointing jittering, the coupling efficiency due to tracking error is given by $\eta_{\text{trk}}(\sigma_\theta) = \exp\!\left(-\frac{\sigma_\theta^{2}}{\theta_d^{2}}\right)$, where $\sigma_\theta$ is the RMS tracking error and $\theta_d = 1.22 \lambda/ D_r$ is the field-of-view of the receiver for the laser wavelength $\lambda$ and receiver aperture diameter $D_r$.

\section{Results \label{sec:res}}
\begin{figure}[ht]
\centering
\includegraphics[width=\linewidth]{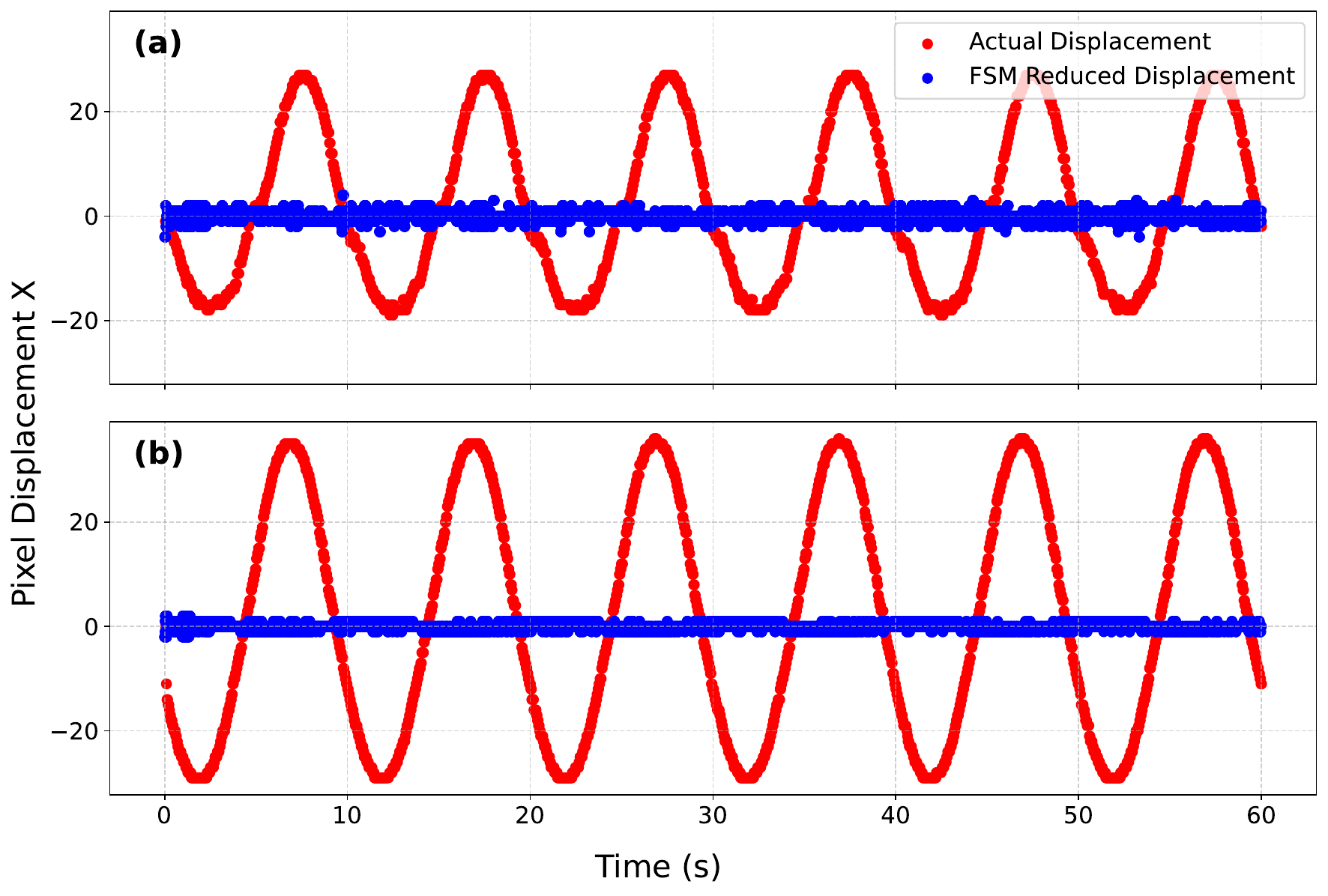}
\caption{FSM compensating the beam displacement for laser input power (a) $0.03\mu W$ and (b) $5.55\mu W$, respectively. Red shows the beam displacement action by the oscillating mirror while blue shows compensation by the FSM.}
\label{fig:comp}
\end{figure}

We have demonstrated the functional purpose of the tracking system by displacing the laser beam by the oscillating mirror, see figure \ref{fig:block}, estimated the camera pixel displacement by the computer and counteracted by the FSM that nullifies the displacement. The first moment of the beacon beam centroid is estimated and used as the input for the FSM. The test results at two laser intensities are given in Fig. \ref{fig:comp}, only $x$ pixel positions are shown.

\begin{figure}[ht]
\centering
\includegraphics[width=\linewidth]{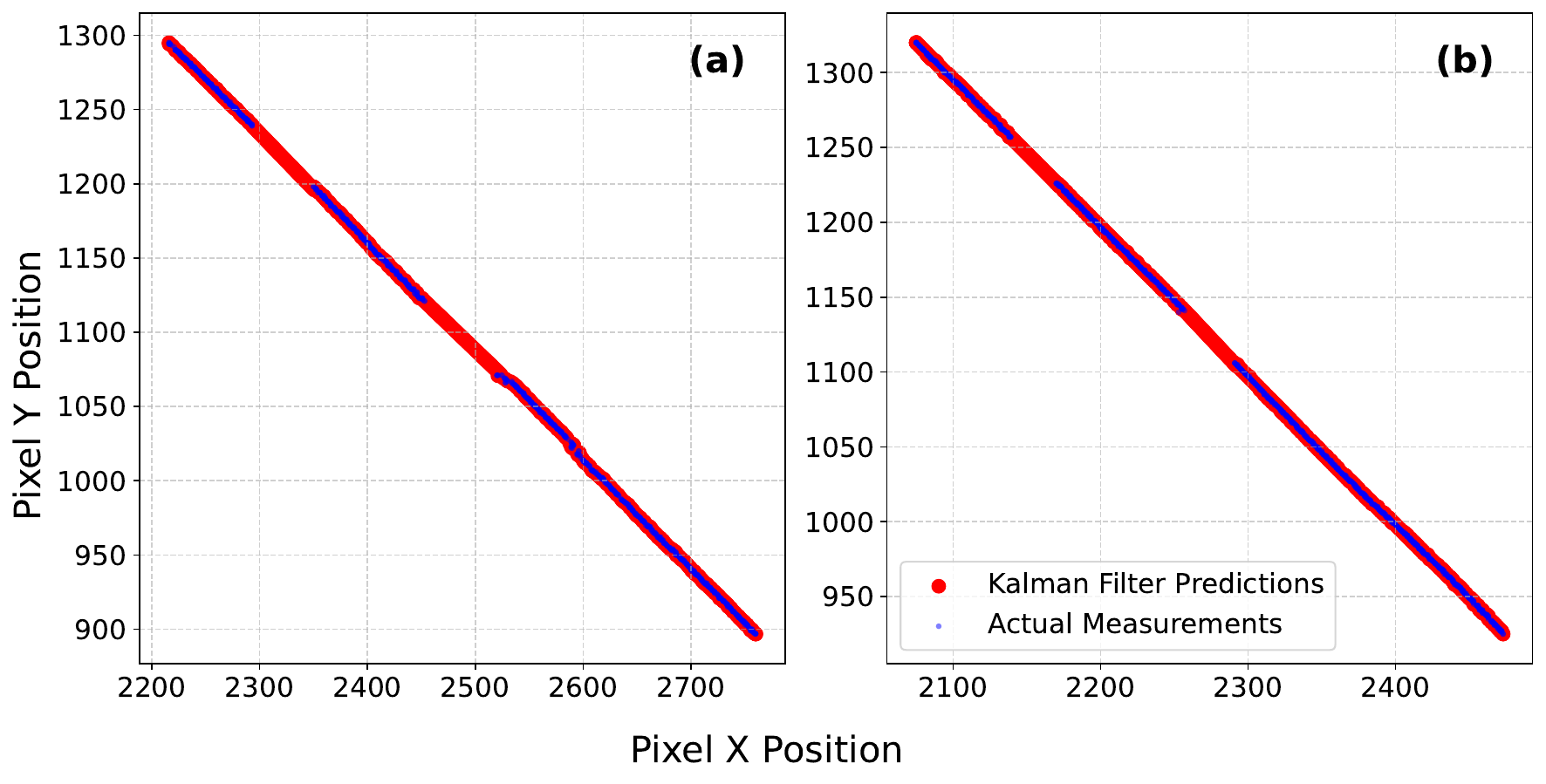}
\caption{$\mathbb{CV}$ model tracking with $34~\text{mW}$ (a) and $5.6~\text{W}$ (b) laser power at the satellite. Pixel $x$ index vs $y$ is shown. The blue line shows the actual laser beam movement across the camera and red line shows the tracking.}
\label{fig:CV_result1}
\end{figure}

For tracking with Kalman filter, initial 100 beam centroid data points reserved for the learning iteration. 
The laser is intermittently blocked to mimic the cloud coverage, in order to test whether the prediction can continue the tracking and rejoin the trajectory after the block is removed. The $\mathbb{CV}$ model tracking result is shown in Fig. \ref{fig:CV_result1} for $5.6~\text{W}$ and for $34~\text{mW}$ laser power at the satellite, corresponds to $5.55~\mu\text{W}$ and $0.03~\mu\text{W}$ at the camera. It shows that the tracking continues in the absence of the laser beacon.

\begin{figure}[ht]
\centering
\includegraphics[scale=0.3]{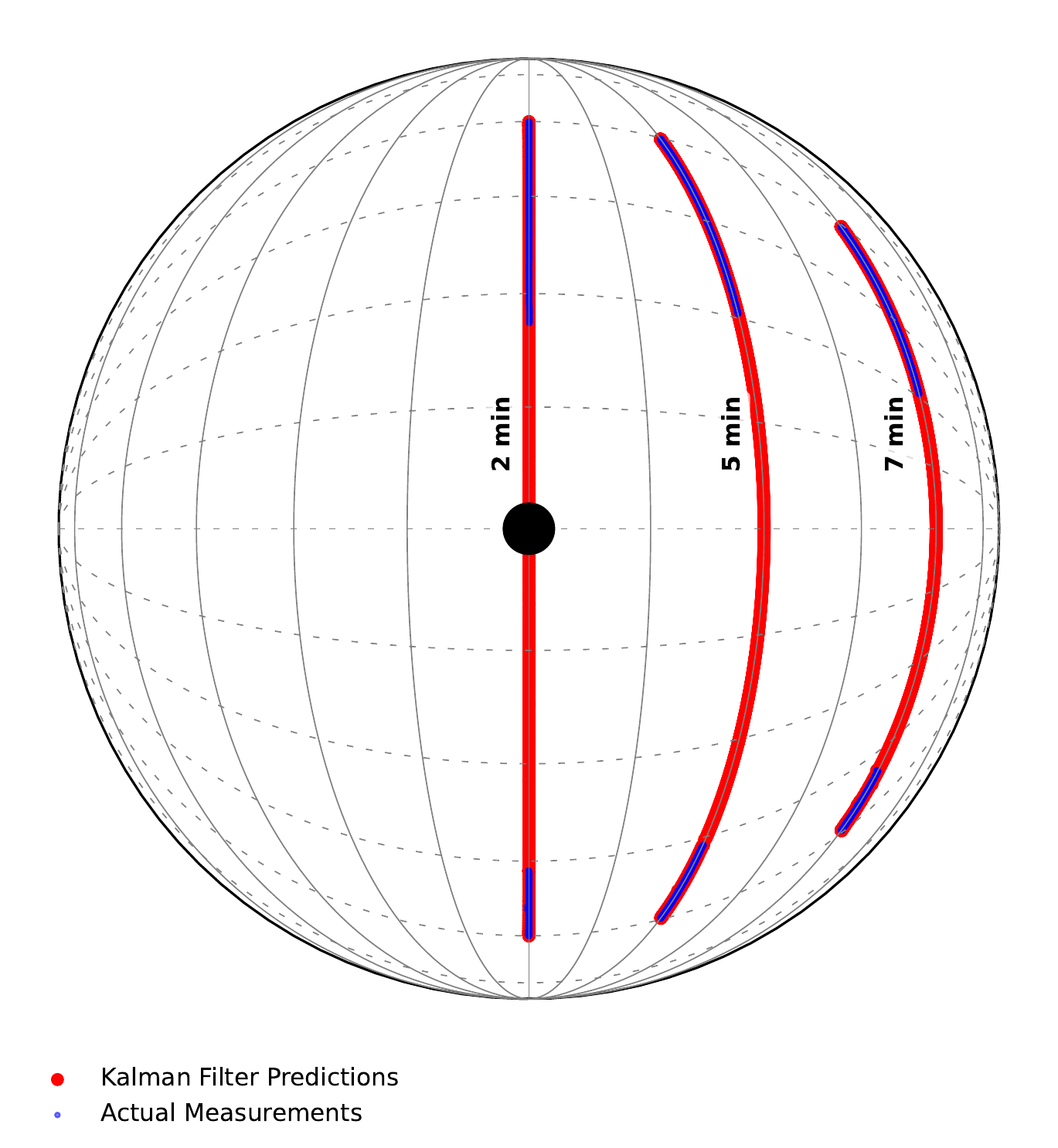}
\caption{Worm's-eye view for CJ model tracking for various satellite trajectories spanning different transition time. The centre black dot represents the blind spot that OGS cannot access.}
\label{fig:CJ_result}
\end{figure}

\begin{figure}[ht]
\centering
\includegraphics[width=\linewidth]{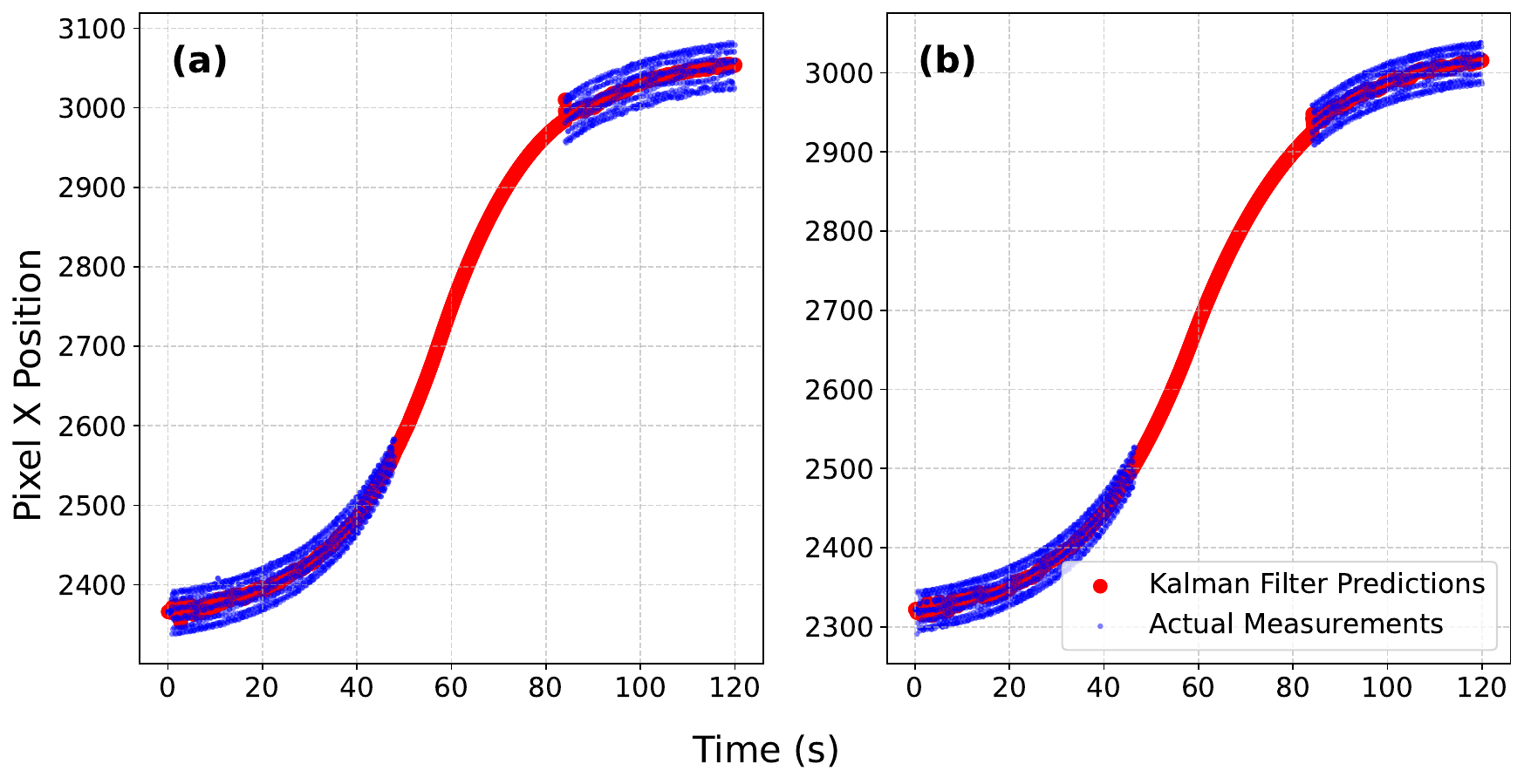}
\caption{ Tracking in the presence of noise under $\mathbb{CJ}$ model for laser input power (a) $0.03\mu W$ and (b) $5.55\mu W$, respectively. Pixel displacement along the $x$ axis against time is shown. Blue is the data with laser on with noise, red is the prediction.}
\label{fig:CJ_result_noise}
\end{figure}

The $\mathbb{CJ}$ model results given in Fig. \ref{fig:CJ_result} correspond to the laser power of $34~\text{mW}$ at the satellite. The $\mathbb{CJ}$ model accounts for the variable acceleration model in which the satellite appeared to accelerate towards the OGS from the horizon to the Zenith angle and then decelerate towards the horizon. To account this, the beam displacement by the FSM is made non-linear in time. The beam is also blocked to mimic obstruction due to cloud coverage. The results for three satellite elevation angles $90^\circ$, $60^\circ$ and $30^\circ$ respectively for 2min, 5min, and 7min transition time, are given in Fig \ref{fig:CJ_result}. Meanwhile, Fig. \ref{fig:CJ_result_noise} shows the tracking in the presence of noise. We can see that the prediction works as expected.

\subsection{Secret key fraction}

\begin{figure}[h]
\centering
\includegraphics[width=0.95\linewidth]{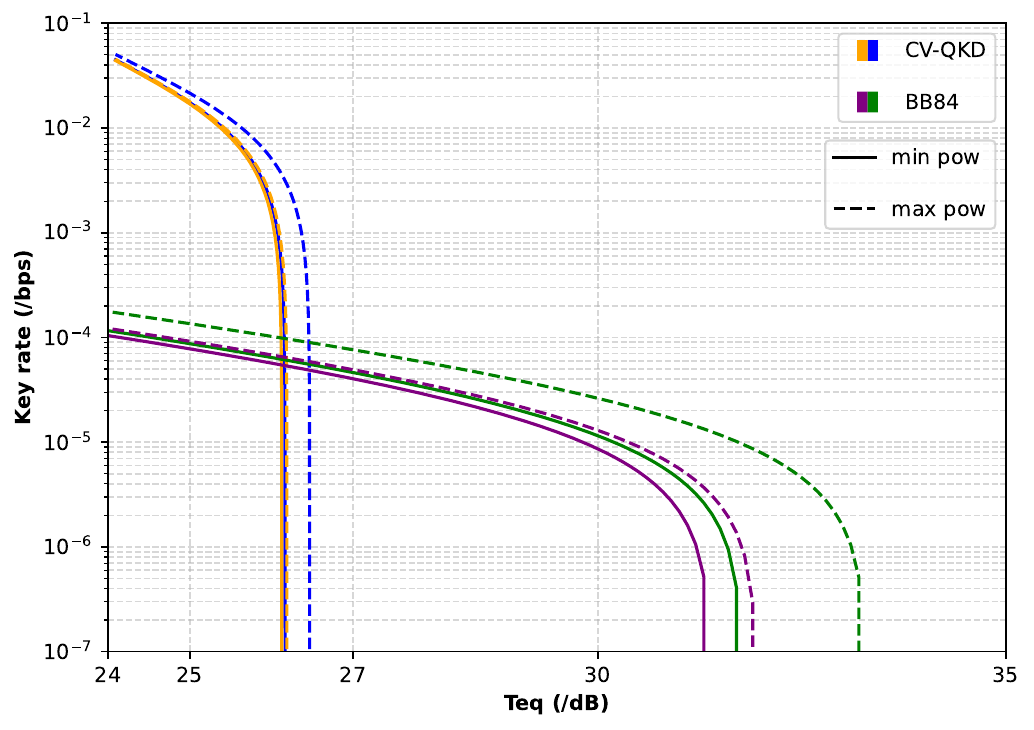}
\caption{The achievable secret key rates as a function of loss for both DV-QKD (green and purple lines) and CV-QKD (blue and orange lines) protocol is plotted here. The coloured solid (dashed) lines indicate the key rate for minimum (maximum) power of beacon laser. The blue \& green (orange \& purple) lines correspond to simulation of a satellite with constant velocity - $\mathbb{CV}$ (continuous acceleration - $\mathbb{CA}$) model.}
\label{fig:SKR}
\end{figure}

We use tracking error as the performance benchmark of our Kalman filer models which also include the ability of our code to correctly identify the laser beam centroid to track the pixel trajectory. We have estimated the RMS error in the tracking in high and low power setting. By using Eq. (\ref{eq:QBER}) for QBER in the case of DV-QKD and using Eq. (\ref{eq:SNR}) for CV-QKD. From the Fig. \ref{fig:SKR}, we can see that the difference are negligible which indicates the effectiveness of out proposal.

Below we plot the graphs for the achievable secret key rates vs loss for both BB84 and CV-QKD protocols in Fig. \ref{fig:SKR}. We assume the channel to have losses as given in App. \ref{app:DV} (BB84) and App. \ref{app:CV} (CV-QKD), with the respective key rates plotted for various tracking errors of low and high power. From the variances in tracking as observed in Fig. \ref{fig:CV_result1} for $\mathbb{CV}$ model (Fig. \ref{fig:CJ_result} for $\mathbb{CJ}$) model, we find the tracking error for low power as $0.27107$ ($0.30525$) and high power as $0.02661$ ($0.25244$) respectively. The losses for quantum signals are typically around $30 - 40~\text{dB}$ for LEO based CubeSats \cite{ghalaii2023satellite, sidhu2021advances} which is different from the losses for beacon laser, which could be upto $60~\text{dB}$ \cite{zhang2021timing, zhang2024modelling}. This is essentially due to the fact that the telescope used for beacon laser has smaller aperture of about $2-5~\text{cm}$, where as that for quantum signals usually have $8-10~\text{cm}$.

\section{Discussions and conclusions \label{sec:conc}}

We consider the aspect of the requirement of high power beacon laser for tracking and pointing, and show that low power is sufficient for active tracking even in the presence of a partial or full blind spots. This is an important result, especially in the context of CubeSats, where total operational power is limited. Thus, given the power required for beacon laser can be minimised, and therefore the accessible power for quantum payload or other operations can be improved. 

The image processing system described in section \ref{subsec:laser_id}, works well across the full range of laser powers considered here. While the algorithm adequately identifies the laser at both extremes, it is noticed and shown in Fig. \ref{fig:cleaning}, that the variance in tracking error at lower power is higher because of the camera noise floor.

The algorithm to centre the laser using the Fast Steering Mirror (FSM) described in section \ref{subsec:fsm_nr} was successful, but we did encounter a hardware limitation; there was latency between movement of the FSM and the respective camera frame representing the movement. At full resolution, this delay caused the system to react to old data, leading to overcorrection. To solve this, we reduced the camera’s region of interest. This allowed the camera to read frames much faster, reducing the delay and increasing accuracy.

This project demonstrates a optical tracking system capable of following a realistic satellite path and an active beam stabilisation system working consistently at all laser power levels, though the minimum power was marginally less stable than the maximum.

Because the Kalman filter is highly dependent on the quality of the initial measurements, any anomalies during this phase can significantly reduce prediction accuracy. These data points are gathered when the satellite is at a low elevation, near 30 degrees, where measurement variance is higher due to atmospheric turbulence and satellite jitter, etc. To counteract these effects, we propose using elevated laser power, instead one near the camera noise floor, in our case $34~\text{mW}$ at the satellite. Also we have not considered other noise sources such as background noise from the atmosphere which elevates the noise floor at the camera. However, in any case, the electrical power requirement can be significantly reduced from a few Watts to a few $100~\text{mW}$ or lower. This is justifying our call for reduced the beacon laser power at the satellite.

We also provide the achievable key rates for the min and max powers in both $\mathbb{CV}$ and $\mathbb{CJ}$ model. This is important considering the fact that the tracking error with the considered feedback even with relatively low power could still work well in high loss regions (Fig. \ref{fig:SKR}). This in turn leads to an interesting aspect of the possibility of even uplink based quantum communication which usually requires power orders of magnitude higher. Our study may provide energy efficient CubeSat models for future quantum communications from space. Our work also find applications in High Altitude Platform Satellites (HAPS) based free space optical communication links as well.

\begin{acknowledgments}
E.V, V.N.R and R.K acknowledge the funding support from EPSRC Quantum Communications Hub (Grant number EP/T001011/1).
\end{acknowledgments}

\bibliography{References}

\appendix

\section{Key rate for BB84 protocol \label{app:DV}}

Pointing error reduces the total effective transmissivity as $T_{\text{eff}}$ according to Eq.~(\ref{eq:eta_eff}), which decreases the gain factors $Q_\mu$ and $Q_1$. Meanwhile, tracking error $e_{\text{trk}}$ increases the QBER $E_\mu$, raising the error-correction cost in Eq.~(\ref{eq:keyrate}). Consequently, increasing either $\sigma_\theta$ or $e_{\text{trk}}$ reduces the key-rate and limits the maximum tolerable loss.

We consider the decoy-state BB84 QKD protocol over a free-space optical channel subject to attenuation and pointing imperfections. The secret key rate is evaluated in the finite-key regime with the GLLP formalism \cite{lo2005decoy}, incorporating detector imperfections, background noise, and tracking-induced errors. The overall channel transmittance is modeled as
\begin{equation}
T_{\text{eff}} = T_{\text{ch}}\, \eta_{\text{trk}}(\sigma_\theta),
\label{eq:eta_eff}
\end{equation}

where $T_{\text{ch}}$ is the channel transmittance (in dB), 
$\sigma_\theta$ is the root-mean-square (RMS) pointing jitter, and $\theta_{\text{div}}$ in $\eta_{\text{trk}}$ is the beam divergence angle. The exponential term represents the reduction in received power due to random beam wandering relative to the beam width at the receiver (difference between telescope center to center of the beam) \cite{dequal2021feasibility}. The source is assumed to emit weak coherent pulses with mean photon number $x \in \{\mu,\nu,{\text{vac}}\}$ and following a Poisson distribution $P_n(x)= (e^{-x} x^n)/n!$. The yield of an $n$-photon pulse, including detector dark counts, is modeled as 
\begin{equation}
Y_n = 1-(1-T_{\text{eff}})^n + p_{\text{dark}}\,[1-(1-T_{\text{eff}})^n],
\end{equation}

where $p_{\text{dark}}$ is the dark count probability per detection window. Considering the decoy state formalism, the overall gain for intensity $x$ is
\begin{equation}
Q_x = \sum_{n=0}^{\infty} P_n(x)\, Y_n.
\end{equation}

However, we truncate the sum at $n=N_{\max}$ with negligible error, for analytical solution. The error rate for an $n$-photon pulse is then given by
\begin{equation}
e_n = \frac{(e_{\text{det}}+e_{\text{trk}})\,[1-(1-T_{\text{eff}})^n] + p_{\text{dark}}/2}{Y_n},
\end{equation}

where $e_{\text{det}}$ is the intrinsic detector misalignment error and $e_{\text{trk}}$ represents polarization or basis errors induced by imperfect tracking. The overall QBER thus becomes
\begin{equation}
E_x = \frac{\sum_{n=0}^{\infty} P_n(x)\, Y_n\, e_n}{Q_x}.
\label{eq:QBER}
\end{equation}

The single-photon yield and single-photon error rate are estimated as $Y_1 = Y_{n=1}, ~~ e_1 = e_{n=1}$, and the single-photon gain is $Q_1 = \mu e^{-\mu} Y_1$. Under the standard decoy state BB84 analysis, the finite-key secret key rate per pulse is

\begin{equation}
R \ge q\!\left[ Q_1\big(1-H_2(e_1)\big) - Q_\mu f_{\text{EC}} H_2(E_\mu) - \Delta_{\text{FK}} \right],
\label{eq:keyrate}
\end{equation}

where $q=1$ is the basis-sifting factor, $f_{\text{EC}}$ is the error-correction inefficiency, and $H_2(x)=-x\log_2 x-(1-x)\log_2(1-x)$ is the binary entropy function, and the finite-key correction term is

\begin{equation}
\Delta_{\text{FK}} = \sqrt{\frac{\ln(2/\varepsilon_{\text{PA}})} {2N\,p_\mu Q_\mu}},
\end{equation}

where $N$ is the total number of transmitted pulses, $p_\mu$ is the probability of sending a signal state, and $\varepsilon_{\text{PA}}$ is the privacy-amplification security parameter. The final secret key rate in bits per second can be obtained by multiplying $R$ by the clock rate $f_{\text{clk}}$.

\section{Key rate for CV-QKD protocol \label{app:CV}}

In the case of CV-QKD, the tracking error increases both the excess noise as well as the effective channel transmittance $T_{\text{eff}}$. The dependency of SNR to the tracking error is,
\begin{equation}
\text{SNR}(\sigma_\theta) = \frac{T_{\text{eff}}\,V_A}{1+\chi_{\text{tot}}},
\label{eq:SNR}
\end{equation}

with
\begin{equation}
\chi_{\text{tot}} = \frac{1-T_{\text{eff}}}{T_{\text{eff}}}
+ \xi + \frac{1}{T_{\text{eff}}} 
\left( \frac{1-\eta_d}{\eta_d} + \frac{v_{\text{el}}}{\eta_d} \right),
\end{equation}

where $\xi$ is the excess noise at Alice, $\eta_d$ is the homodyne detector efficiency, and $v_{\text{el}}$ is the electronic noise and $V_A$ is the signal modulation variance. For the asymptotic regime in CV-QKD, considering reverse reconciliation, the secret key rate is given by \cite{laudenbach2018continuous},
\begin{equation}
K=\beta I_{AB} - \chi_{BE}.
\end{equation}

Here $\beta$ is the reconciliation efficiency, $I_{AB}$ is the mutual information between Alice and Bob and $\chi_{BE}$ is the Holevo information between Bob and Eve. The mutual information in the conventional way with a homodyne detection scheme is given by,
\begin{equation}
I_{AB}= \frac{1}{2}\log_{2}\frac{T_{\text{eff}}V+\chi_{\text{tot}}}{1+\chi_{\text{tot}}},
\label{eq:iab}
\end{equation}

where $V$ is the modulation variance, and $\chi_{\text{tot}}$ is the total noise at Alice. Here we find the covariance matrix and Holevo bound for the bypass channel under the assumption of restricted Eve \cite{ghalaii2023satellite}. The protocol is mapped to an entanglement based one (with two mode squeezed vacuum state), to find the respective covariance matrices as well as Holevo information (reverse reconcilation), which is given by
\begin{equation}
\chi_{BE}=H(EE^{\prime})-H(EE^{\prime}|B).
\end{equation}

The conditional entropies $H(EE^{\prime})$ and $H(EE^{\prime}|B)$ can be obtained by the corresponding symplectic eigenvalues of the covariance matrix. Given $\Lambda_{1}$ \& $\Lambda_{2}$ are the symplectic eigenvalues corresponding to Eve's variance $V_{EE^{\prime}}$, and $\Lambda_{3}$ \& $\Lambda_{4}$ to $V_{EE^{\prime}|B}$, we find
\begin{equation}
\chi_{BE}=g(\Lambda_{1})+g(\Lambda_{2})-g(\Lambda_{3})-g(\Lambda_{4}),
\end{equation}
where $g(x)=(\frac{x+1}{2})\log_{2}(\frac{x+1}{2})-(\frac{x+1}{2})-\log_{2}(\frac{x+1}{2})$.

For a detailed description of the same, refer \cite{medlock2026continuous}.

\end{document}